\documentclass[conference]{IEEEtran}
\IEEEoverridecommandlockouts
\usepackage{cite}
\usepackage{amsmath,amssymb,amsfonts}
\usepackage{graphicx}
\usepackage{textcomp}
\usepackage{xcolor}
\usepackage[a4paper, total={184mm,239mm}]{geometry}
\def\BibTeX{{\rm B\kern-.05em{\sc i\kern-.025em b}\kern-.08em
    T\kern-.1667em\lower.7ex\hbox{E}\kern-.125emX}}

\usepackage{booktabs} 
\usepackage{gensymb}
\usepackage[english]{babel}
\usepackage{bm}
\usepackage[capitalise]{cleveref}
\usepackage{subcaption}
\usepackage{pifont}
\usepackage{makecell}
\usepackage{multirow}
\usepackage{multicol}
\usepackage{algorithm}
\usepackage{algpseudocode}

\newcommand{\Vmin}{V_{min}}

\newcommand{\loss}{\mathcal{L}}

\newcommand{\x}{\bm{\mathrm{x}}}
\newcommand{\w}{\bm{\mathrm{w}}}
\newcommand{\y}{\mathrm{y}}
\newcommand{\z}{\bm{\mathrm{z}}}
\newcommand{\res}{\mathrm{r}}
\newcommand{\bias}{\mathrm{b}}
\newcommand{\X}{\bm{\mathrm{X}}}
\newcommand{\Y}{\bm{\mathrm{y}}}

\newcommand{\R}{\mathbb{R}}

\DeclareMathOperator*{\argmin}{arg\,min}

\makeatletter
\newcommand{\smallbullet}{} 
\DeclareRobustCommand\smallbullet{%
  \mathord{\mathpalette\smallbullet@{0.7}}%
}
\newcommand{\smallbullet@}[2]{%
  \vcenter{\hbox{\scalebox{#2}{$\m@th#1\bullet$}}}%
}
\makeatother

\begin{document}

\title{Data Efficient Prediction of Minimum Operating Voltage via Inter- and Intra-Wafer Variation Alignment
}

\author{
\IEEEauthorblockN{Yuxuan Yin}
\IEEEauthorblockA{\textit{Electrical and Computer Engineering} \\
\textit{University of California}\\
Santa Barbara, CA, USA \\
y\_yin@ucsb.edu}
\and
\IEEEauthorblockN{Rebecca Chen}
\IEEEauthorblockA{\textit{Automotive Processing} \\
\textit{NXP Semiconductors} \\
Austin, TX, USA \\
rebecca.chen\_1@nxp.com}
\and
\IEEEauthorblockN{Chen He}
\IEEEauthorblockA{\textit{Automotive Processing} \\
\textit{NXP Semiconductors} \\
Austin, TX, USA \\
chen.he@nxp.com}
\and
\IEEEauthorblockN{Peng Li}
\IEEEauthorblockA{\textit{Electrical and Computer Engineering} \\
\textit{University of California}\\
Santa Barbara, CA, USA \\
lip@ucsb.edu}
}

\maketitle

\begin{abstract}
Predicting the minimum operating voltage ($V_{min}$) of chips stands as a crucial technique in enhancing the speed and reliability of manufacturing testing flow. However, existing $V_{min}$ prediction methods often overlook various sources of variations in both training and deployment phases. Notably, the neglect of wafer zone-to-zone (intra-wafer) variations and wafer-to-wafer (inter-wafer) variations, compounded by process variations, diminishes the accuracy, data efficiency, and reliability of $V_{min}$ predictors. To address this gap, we introduce a novel data-efficient $V_{min}$ prediction flow, termed restricted bias alignment (RBA), which incorporates a novel variation alignment technique. Our approach concurrently estimates inter- and intra-wafer variations. Furthermore, we propose utilizing class probe data to model inter-wafer variations for the first time. We empirically demonstrate RBA's effectiveness and data efficiency on an industrial 16nm automotive chip dataset.
\end{abstract}


\begin{figure*}[htbp]
     \centering
     \begin{subfigure}[b]{0.48\textwidth}
         \centering
         \includegraphics[width=\textwidth]{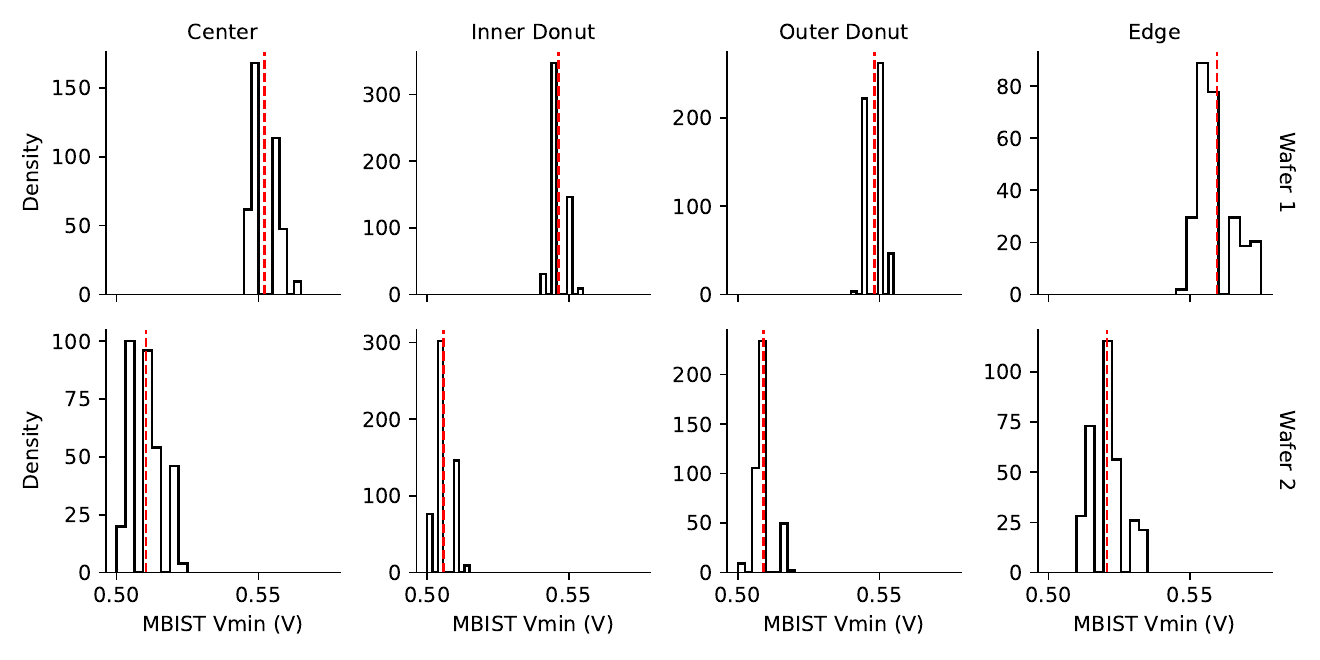}
         \caption{MBIST $\Vmin$ variation}
         \label{fig:vmin-variation}
     \end{subfigure}
     \hfill
     \begin{subfigure}[b]{0.48\textwidth}
         \centering
         \includegraphics[width=\textwidth]{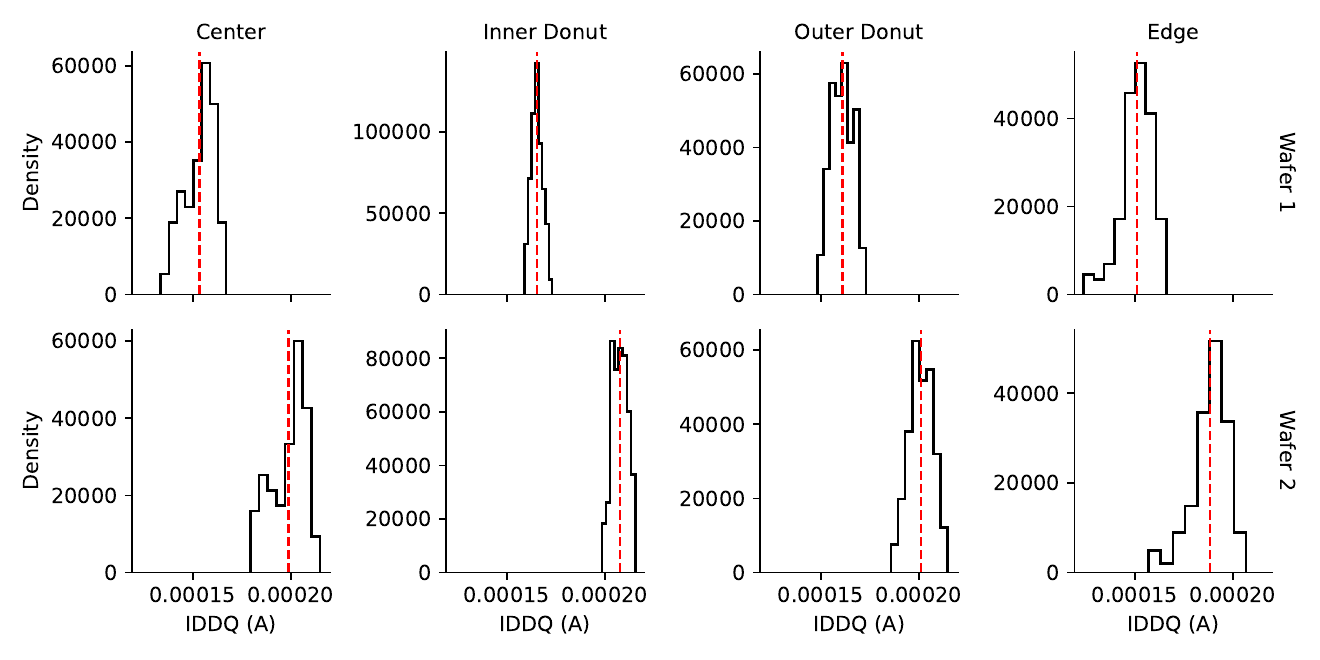}
         \caption{IDDQ current variation}
         \label{fig:feature-variation}
     \end{subfigure}
        \caption{Wafer-to-wafer and wafer zone variations across 2 wafers, measured at 25\degree C. Red dashed lines represent mean values.}
        \label{fig:wafer-variation}
\end{figure*}

\section{Introduction}
The measurement of the minimum operating voltage ($V_{min}$) represents a pivotal testing procedure crucial for assessing chip performance. It enables the identification of substandard products, facilitates power consumption optimization, and serves as an early indicator of potential failures during the device's lifespan. A case study involving 7nm industry chips illustrates that subjecting all chips to uniform energy levels leads to a minimum 16\% increase in energy utilization \cite{VminBin}.

As technology nodes continue to shrink, the significance of $\Vmin$ tests employing structural test patterns (e.g., SCAN) amplifies, becoming indispensable for pinpointing minute flaws and defects \cite{VminTest} within chips. For instance, in Fin Field-Effect Transistor (FinFET) technology, a resistive short defect within a critical path may augment leakage current, potentially evading detection through IDDQ testing due to high background leakage. However, such defects can manifest as discernible $\Vmin$ degradation when subjected to $\Vmin$ testing (e.g., SCAN $\Vmin$) traversing the affected critical path.


\begin{figure*}[tbp]
    \centering
    \includegraphics[width=0.95\textwidth]{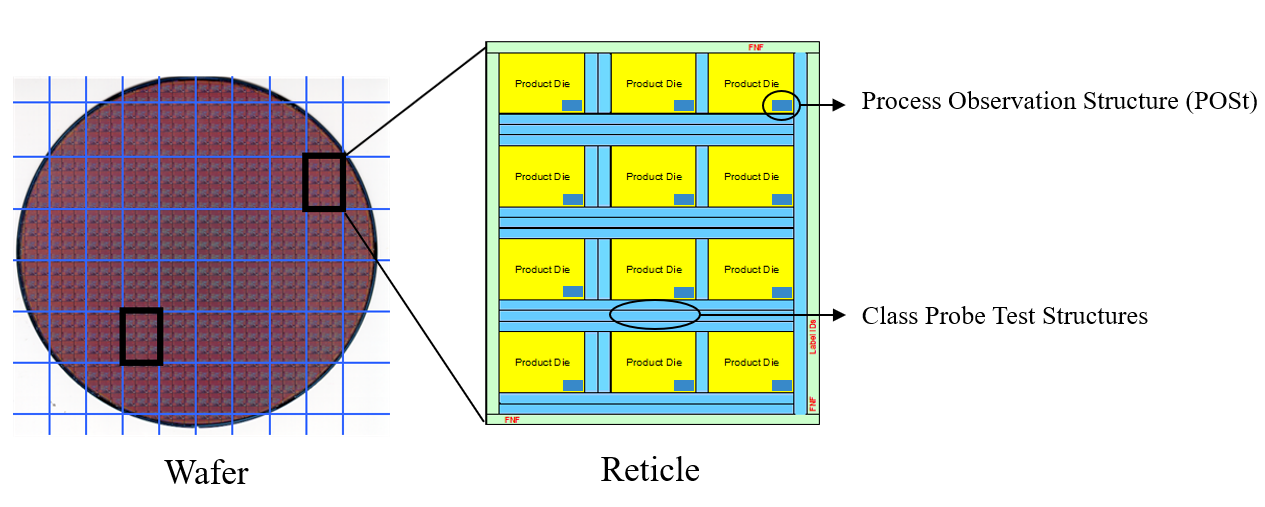}
    \caption{Class probe test structures and process observation structures}
    \label{fig:class-probe}
\end{figure*}

Testing $\Vmin$ voltage incrementally for each die during manufacturing is impractical due to its time-consuming nature. Instead, a pragmatic approach involves testing $\Vmin$ within a predefined target range, rather than exhaustively covering the entire testing space. However, owing to process variations, chip properties exhibit variability from die to die, wafer zone to wafer zone, and wafer to wafer. The presence of such variations precludes the establishment of a fixed target $\Vmin$, as doing so would lead to significant underkill for failures and overkill for normal dies.

Current industrial practices rely on die-level features to construct $\Vmin$ prediction models that account for die-to-die variations. These features, gathered from parametric tests or on-chip monitors such as IDDQ tests and ring oscillators \cite{Odometer, RODesign}, serve as inputs to machine learning-based $\Vmin$ predictors \cite{Accumulative, SVMFmaxBinning, GPModel, ML-Assisted, yin-mln-itc}. However, existing methodologies fall short in capturing wafer zone-to-zone (intra-wafer) and wafer-to-wafer (inter-wafer) variations. \cref{fig:vmin-variation} and \cref{fig:feature-variation} illustrate the impact of inter- and intra-wafer variations to $\Vmin$ and parametric features in an industrial 16nm chip dataset, respectively. It is evident that either type of process variation significantly alters the distribution of $\Vmin$ and parametric features, ultimately impairing the accuracy of aforementioned $\Vmin$ predictors. 

In this paper, we introduce a novel $\Vmin$ prediction framework called restricted bias alignment (RBA), designed to systematically capture inter- and intra-wafer variations, along with die-to-die variations. To address die-level variations, we adopt parametric test features in line with prior research.  However, for inter- and intra-wafer variations, we treat them as independent and employ a voltage bias term to model their respective impacts on individual dies. Additionally, we harness class probe data to model inter-wafer variations.  By aligning and modeling process variations, RBA is data-efficient and robust when training, and is accurate in deployment for dies from new wafers. Our main contributions are:

    $\smallbullet$ We propose a novel data-efficient algorithm for estimating and aligning $\Vmin$ shifts resulting from inter- and intra-wafer variations.
    
    $\smallbullet$ We propose to utilize class probe data for inter-wafer $\Vmin$ shift modeling for the first time, and propose to reuse pre-learned intra-wafer $\Vmin$ shift for dies from new wafers in addressing process variations.

    $\smallbullet$ Through empirical analysis, we demonstrate the effectiveness and data efficiency of the proposed $\Vmin$ prediction approach on an industrial dataset.

\begin{figure}[tb]
    \centering
    \includegraphics[width=0.48\textwidth]{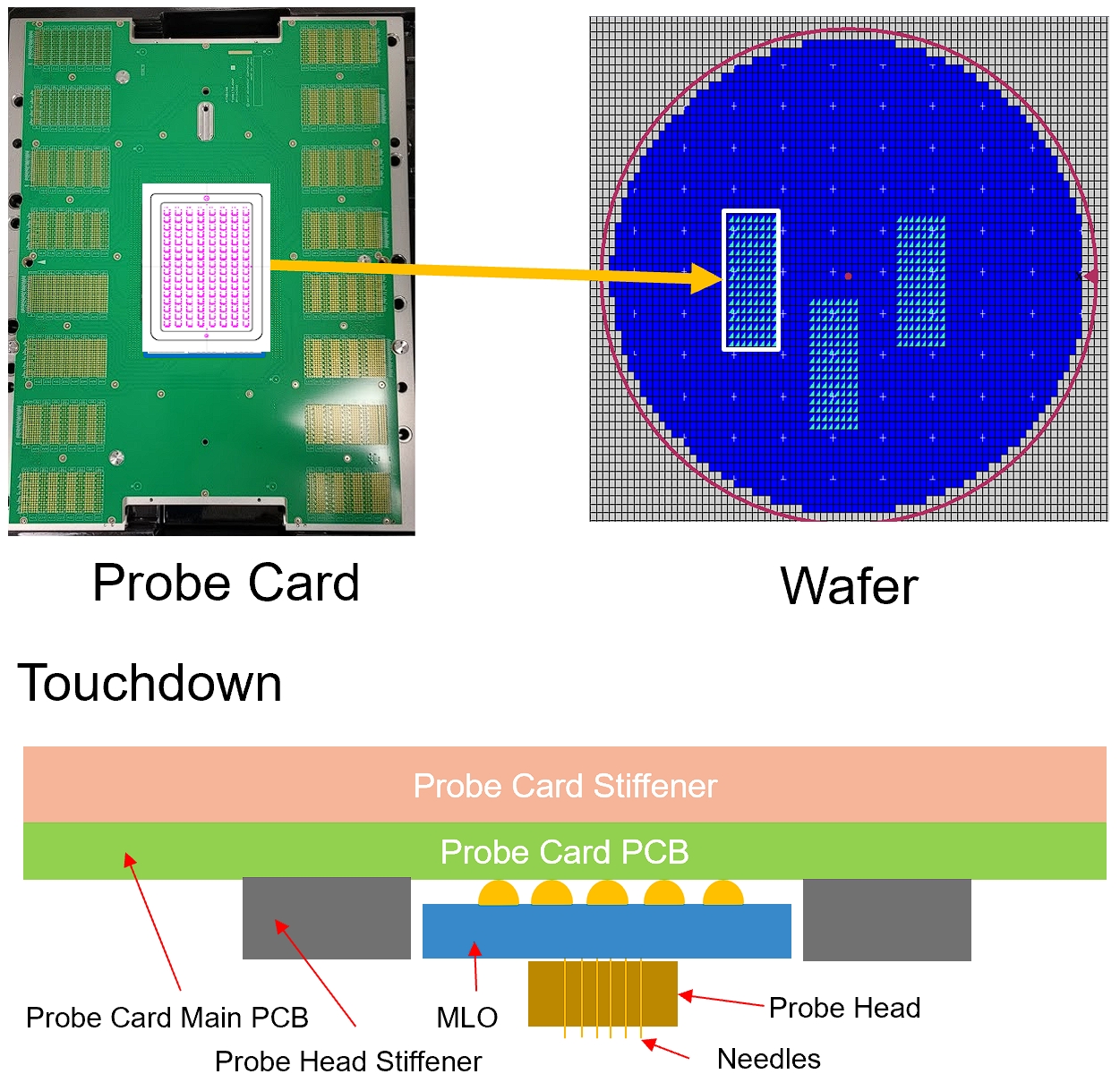}
    \caption{A highly-parallel wafer probe touchdown}
    \label{fig:touch-down}
\end{figure}
\begin{figure}[!tbp]
    \centering
    \includegraphics[width=0.43\textwidth]{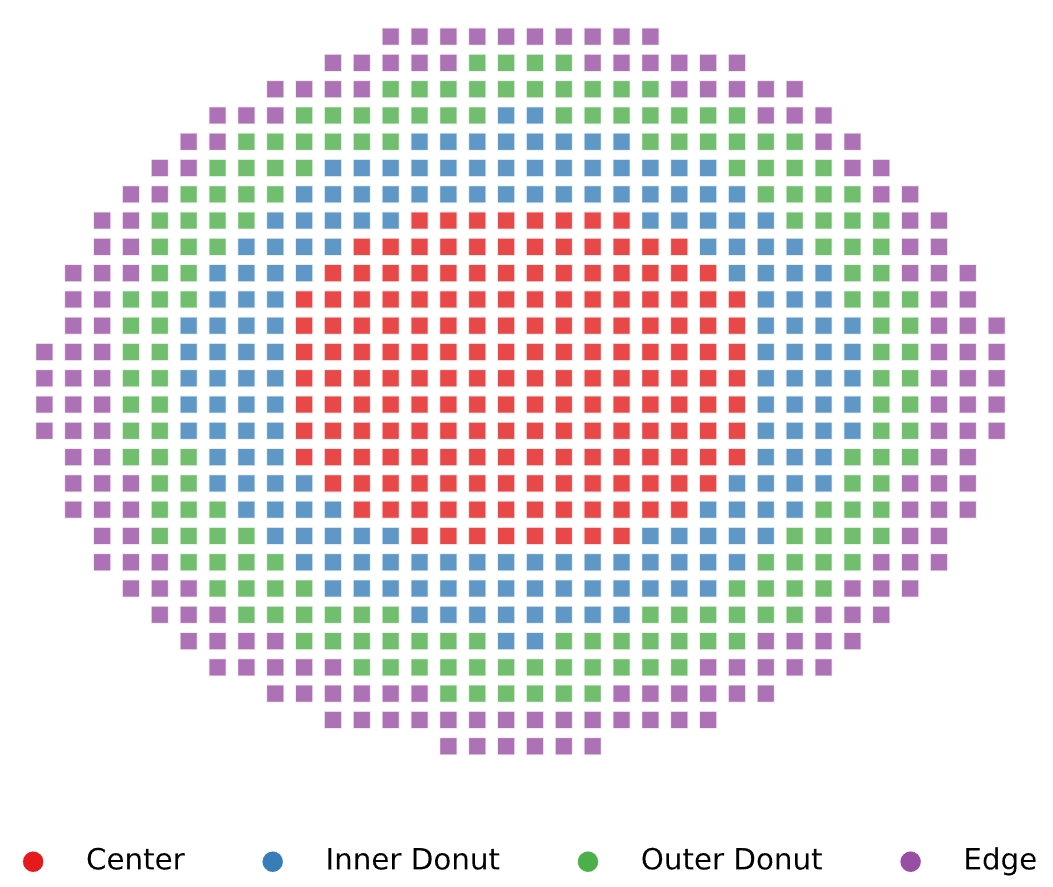}
    \caption{A balanced wafer region partition into 4 regions}
    \label{fig:wafer-region}
\end{figure}

\section{Preliminaries}
\subsection{Testing Flow in Semiconductor Manufacturing}
In semiconductor manufacturing, wafer-level testing employs class probe test structures that are situated in the scribe lines between product dies, outside the actual chips, as depicted in \cref{fig:class-probe}. These test structures typically contain components like transistors, via chains, resistors, and capacitors, mirroring the fabrication process used for the product dies. The purpose of these test structures is to provide feedback on the wafer's processing, enable statistical process control, and help reduce variations from wafer to wafer.

At the die level, each part incorporates its own set of test structures, known as Process Observation Structures (POSt), typically located at the corners of each die. These structures include components like ring oscillators, transistors, bipolar junctions, resistors, and capacitors, which reflect the elements used in the actual circuits on each die. POSt structures offer visibility into the die-level processing, allowing engineers to monitor and evaluate the performance of individual dies. \cref{fig:touch-down} shows an example of a highly parallel probe hardware system. The probe card touches down on multiple Devices Under Test (DUTs) on the wafer to drive electrical stimulus to all dies being tested. Following a predefined algorithm, the probe card moves around to cover the entire wafers. 

Together, these two sets of test structures—the class probe at the wafer level and the POSt at the die level—create a hierarchical system for tracking process variations across wafers and individual dies. This setup allows for correlations between class probe data and inter-wafer Vmin bias, as well as between POSt data and the Vmin of individual dies. The presence of these correlations indicates that the test structures can be used to predict and control process variations, contributing to more reliable and consistent chip production.


\subsection{Linear Regression for $\Vmin$ Prediction} \label{sec:prlm-linear-regression}
Linear regression is a simple yet effective method to predict $\Vmin$. It builds upon the following assumption
\begin{equation}
    \y = \x \w + \bias+ \epsilon
\end{equation}
where $\y \in (0,+\infty)$ is the positive value of $\Vmin$, $x \in \R^{1\times d}$ is a $d$-dimensional row vector, which is a subset of features measured by parametric tests, $\w\in \R^{d\times1}$ is a $d$-dimensional column vector of unknown parameters, $\bias\in \R$ is a bias term of $\Vmin$, and $\epsilon \in \R$ accounts for the influence on $\Vmin$ other than features $\x$.

Given a training dataset $(\X,\Y)$, one can estimate $\widehat{\w}$ and $\widehat{\bias}$ via minimizing the sum of square residuals
\begin{equation}
    \widehat{\w}, \widehat{\bias} = \argmin_{\w,\bias} ||\Y-\X \w-\bias||_2^2
\end{equation}
and the solution is 
\begin{align}
    \widehat{\w} & = \big(\widetilde{\X}^T \widetilde{\X} \big) ^{-1} \widetilde{\X}^T \widetilde{\Y} \\
    \widehat{\bias} & = \bar{\Y} - \bar{\X} \widehat{\w}
\end{align}
where the bar operator $\bar{\cdot}$ computes the mean value (vector) of a vector (matrix), and the tilde operator $\widetilde{\cdot}$ centralizes the input.

The learned parameters $\widehat{\w}$ and $\widehat{\bias}$ are used to predict $\Vmin$ of new dies. For a new chip, we perform parametric tests to collect the input feature $\x^{test}$ for linear regression, and then compute the $\Vmin$ prediction $\widehat{\y}^{test}$:
\begin{equation}
    \widehat{\y}^{test} = \x^{test} \widehat{\w} + \widehat{\bias}
\end{equation}

\begin{figure}
    \centering
    \includegraphics[width=0.48\textwidth]{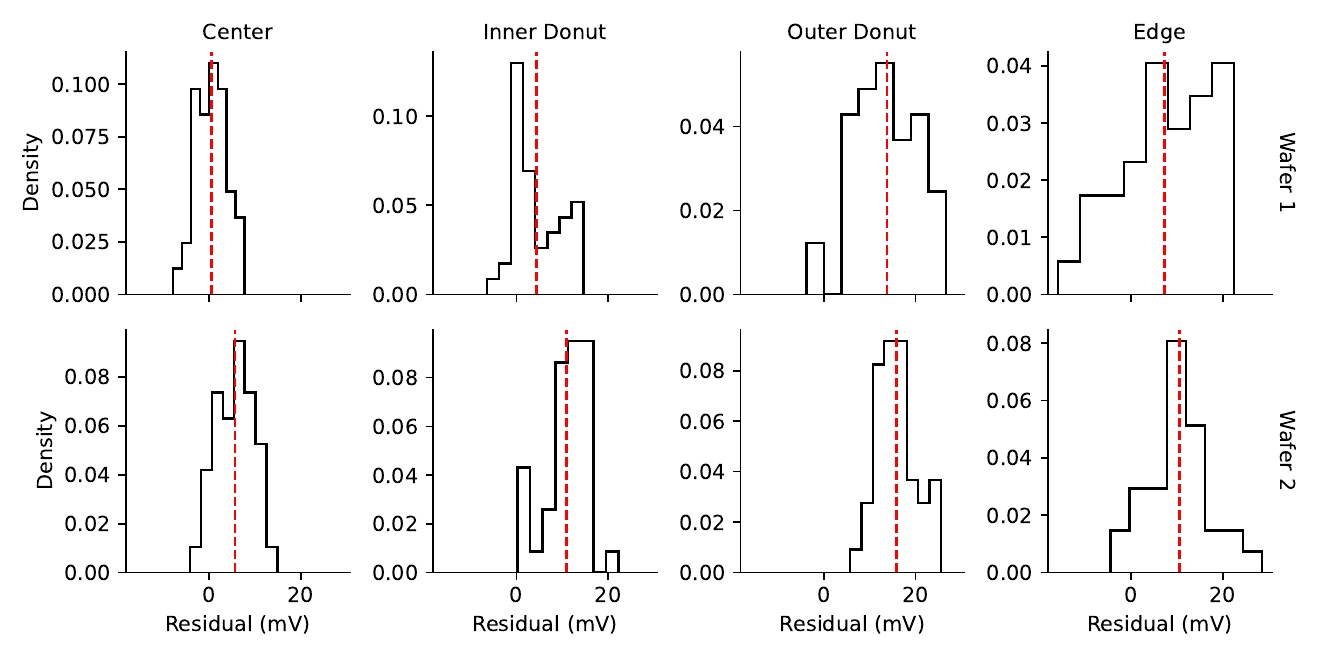}
         \caption{Variation of testing residuals of $\Vmin$ prediction of a linear regression trained on dies from the center zone from wafer 1. Red dashed lines represent mean values.}
         \label{fig:residual-variation}
\end{figure}

\subsection{Influence of Inter- and Intra-Wafer Variation on $\Vmin$ Prediction}
Process variations are inherent in modern semiconductor manufacturing, with their significance magnifying as technology nodes and wafer sizes scale. Typically, there are two types of process variations: inter-wafer (wafer-to-wafer) variations and intra-wafer (zone-to-zone) variations. 

We visually depict the contributions of both variations to the distribution of $\Vmin$ and parametric features within our industrial 16nm automotive dataset in \cref{fig:wafer-variation}. Specifically, we present histogram plots and mean values of MBIST $\Vmin$ in \cref{fig:vmin-variation} and IDDQ current in \cref{fig:feature-variation}, spanning the four regions of two wafers from the same lot. Each row in either sub-figure represents the intra-wafer variation of a given wafer, while each column signifies the inter-wafer variations of a specific wafer zone.

It is evident that both $\Vmin$ and parametric feature distributions exhibit considerable variance across wafers and regions. However, if process variations merely introduce \textit{covariate shift}, wherein the relationship $\y|\x$ between $\Vmin$ and parametric features remains constant, we could feasibly train a $\Vmin$ predictor and deploy it on new testing dies. Unfortunately, the assumption of covariate shift does not hold for the $\Vmin$ prediction task. To illustrate, we train the aforementioned linear model on dies from the center zone of wafer 1 and test it across all four zones of wafers 1 and 2. The resulting residual of MBIST $\Vmin$ on the testing data is depicted in \cref{fig:residual-variation}, where the residual $\res$ is computed as
\begin{equation}
    \res = \y - \widehat{\y}
\end{equation}

The predictor performs admirably on the center zone of wafer 1; however, its accuracy notably declines on other testing wafer zones of both wafer 1 and wafer 2. This outcome underscores that process variations alter the statistical correlation between $\Vmin$ and parametric test features, rather than solely inducing covariate shift. Consequently, there arises a necessity to systematically address process variations, encompassing both inter- and intra-wafer variations, to attain accurate and robust $\Vmin$ prediction.


\section{Data Efficient Inter- and Intra-Wafer Variation Alignment}
It is clear that both $\Vmin$ and parametric features exhibit significant variation from wafer to wafer and from region to region. In this paper, we concentrate on a specific impact of process variations on the dependency $\y|\x$ between $\Vmin$ and parameter features: a consistent voltage shift of $\Vmin$ relative to the bias $\bias$.

We introduce two methods to align the $\Vmin$ shift resulting from process variations. The first one, named Bias Alignment (BA), is a general machine learning method: it aims to estimate the voltage shift term of each wafer zone The second method, known as Restricted Bias Alignment (RBA), is guided by domain knowledge: it assumes independence between inter- and intra-wafer variations, while the intra-wafer variation remains constant across all wafers. Additionally, for RBA, we propose utilizing class probe features to correlate the estimated inter-wafer $\Vmin$ shift. This technique empowers RBA to predict $\Vmin$ for dies from a new wafer without necessitating access to any training data from that wafer.

\subsection{Bias Alignment (BA) for $\Vmin$ Prediction}
\subsubsection{Problem Formulation}
Assume we have $N$ wafers in the training dataset. Each wafer is partitioned into $M$ balanced wafer zones. We denote $(\X_{i,j}, \Y_{i,j})$ as the collected (feature, $\Vmin$) pair of a batch of dies from the $j$-th zone of the $i$-th wafer. Here $i=1,\cdots,N$ and $j=1,\cdots, M$. Taking process variations into consideration, the $\Vmin$ is modeled as 
\begin{equation}
    \Y_{i,j} = \X_{i,j} \w + \bias_{i,j} + \epsilon
\end{equation}
where $\w$ is the fixed coefficients across wafer zones, and $\bias_{i,j}$ is a bias term of the $i$-th wafer's $j$-th zone, accounting for the impact of both inter- and intra-wafer variations.

We construct a loss function $\loss_{BA}$ as the sum of square residuals of the 
$\Vmin$ prediction across the whole training set:
\begin{equation}\label{eq:ba-loss}
    \loss_{BA} := \ \sum_{i,j} ||\Y_{i,j}-\X_{i,j} \w-\bias_{i,j}||_2^2
\end{equation}
and minimize it to estimate $\widehat{\w}$ and $\widehat{\bm{\bias}}$
\begin{equation}\label{eq:ba-obj}
    \widehat{\w}, \widehat{\bm{\bias}} = \argmin_{\w,\bm{\bias}}  \loss_{BA}\left(\w, \bm{\bias}\right)
\end{equation}
where $\bm{\bias}$ is a set containing all biases $\bias_{i,j}$.

\subsubsection{Solution} We provide analytical solution of \cref{eq:ba-obj}. For $\widehat{\w}$ and $\widehat{\bm{\bias}}$ we should have 
\begin{align}
    \frac{\partial \loss_{BA}}{\partial \widehat{\w}} & = 0  \label{eq:ba-w} \\
     \frac{\partial \loss_{BA}}{\partial \widehat{\bm{\bias}}} & = 0 \label{eq:ba-bias}
\end{align}
From \cref{eq:ba-bias} we have 
\begin{equation}
    \frac{\partial }{\partial \widehat{\bias}_{i,j}} ||\Y_{i,j}-\X_{i,j} \widehat{\w}-\widehat{\bias}_{i,j}||_2^2 = 0
\end{equation}
which means that 
\begin{equation}\label{eq:ba-bias}
    \widehat{\bias}_{i,j} = \bar{\Y}_{i,j} - \bar{\X}_{i,j} \widehat{\w}
\end{equation}
Then \cref{eq:ba-w} can be formulated as
\begin{equation}
     \frac{\partial }{\partial \widehat{\w}} \sum_{i,j} ||\bar{\Y}_{i,j}-\bar{\X}_{i,j} \widehat{\w}||_2^2 = 0
\end{equation}
where we can directly compute the formula of $\widehat{\w}$:
\begin{equation}\label{eq:ba-w}
    \widehat{\w} = \big(\sum_{i,j} \widetilde{\X}^T_{i,j} \widetilde{\X}_{i,j} \big) ^{-1} \big(\sum_{i,j} \widetilde{\X}^T_{i,j} \widetilde{\Y}_{i,j}\big)
\end{equation}

\subsubsection{Discussion} The estimated coefficient $\widehat{\w}$ and bias $\widehat{\bm{\bias}}$ in BA aggregate the information from dies of all wafer zones, and meanwhile tackle individual $\Vmin$ shift of each wafer zone caused by inter- and intra-wafer variations. Compared with the vanilla linear regression presented in \cref{sec:prlm-linear-regression}, BA is more data-efficient and accurate for $\Vmin$ prediction.

For a testing die $(\x^{test}_{i,j}, \y^{test}_{i,j})$ form the $j$-th zone of the $i$-th wafer, the $\Vmin$ prediction of BA is 
\begin{equation}
    \widehat{\y}^{test} = \x^{test} \widehat{\w} + \widehat{\bias}_{i,j}
\end{equation}

However, in the $\Vmin$ inference process, BA necessitates the computation of the voltage shift term in advance. Consequently, testing $\Vmin$ for several dies from a new wafer (wafer zone) remains necessary. This inherent limitation inspires our second $\Vmin$ prediction approach, wherein no $\Vmin$ test is required for the new wafer (wafer zone).
\begin{table}[]
    \centering
    \begin{tabular}{c c c c c}
    \toprule
    Zone & Center & Inner Donut & Outer Donut & Edge \\
    \midrule
      \# Dies   &  180 & 184 & 184 & 183 \\
      \bottomrule
    \end{tabular}
    \caption{The number of dies in each wafer zone}
    \label{tab:wafer-zone}
\end{table}

\begin{table*}[]
    \centering
\caption{The testing RMSE (mV) of DC Scan $\Vmin$ prediction with 75\% data for training}
    \label{tab:main-rslt-dc-scan}
\begin{tabular}{@{}cl|cccc|cccc@{}}
\toprule
\multicolumn{2}{c|}{Temperature}& \multicolumn{4}{c|}{-45\degree C} & \multicolumn{4}{c}{25\degree C} \\ 
Wafer ID & Wafer Zone       & \# Die & Linear Regression  & Bias Alignment   & Restricted BA  & \# Die & Linear Regression  & Bias Alignment   & Restricted BA  \\ \midrule
1      & Center      & 166    & 3.54 & 2.69 & 2.88 & 165    & 3.28 & 2.72 & 2.68 \\
1      & Inner Donut & 162    & 5.35 & 3.71 & 4.14 & 161    & 5.03 & 4.12 & 4.64 \\
1      & Outer Donut & 158    & 7.69 & 3.24 & 3.28 & 173    & 4.79 & 3.92 & 3.91 \\
1      & Edge        & 91     & 5.21 & 3.34 & 3.05 & 128    & 5.17 & 4.82 & 4.78 \\ \midrule
2      & Center      & 152    & 11.84& 3.80 & 4.90 & 166    & 6.13 & 3.63 & 3.98 \\
2      & Inner Donut & 153    & 8.23 & 3.57 & 3.85 & 172    & 5.16 & 3.62 & 3.79 \\
2      & Outer Donut & 155    & 5.04 & 3.25 & 3.59 & 172    & 4.37 & 4.03 & 3.89 \\
2      & Edge        & 93     & 9.39 & 5.60 & 5.58 & 128    & 7.82 & 5.51 & 5.72 \\ \midrule
3      & Center      & 54     & 8.03 & 4.46 & 4.37 & 152    & 5.49 & 4.36 & 4.38 \\
3      & Inner Donut & 111    & 5.19 & 3.05 & 3.49 & 168    & 3.98 & 3.06 & 3.46 \\
3      & Outer Donut & 126    & 5.74 & 4.52 & 4.69 & 173    & 4.50 & 4.43 & 4.72 \\
3      & Edge        & 89     & 4.03 & 4.07 & 4.27 & 130    & 4.66 & 4.68 & 4.71 \\ \midrule
4      & Center      & 138    & 4.25 & 3.73 & 3.60 & 161    & 3.58 & 2.43 & 2.38 \\
4      & Inner Donut & 157    & 3.98 & 2.72 & 2.72 & 174    & 3.63 & 3.32 & 3.46 \\
4      & Outer Donut & 154    & 5.16 & 2.92 & 2.94 & 171    & 5.61 & 3.93 & 3.88 \\
4      & Edge        & 91     & 3.52 & 3.56 & 3.55 & 128    & 5.18 & 4.58 & 4.88 \\ \midrule
5      & Center      & 153    & 5.98 & 3.12 & 3.49 & 171    & 3.80 & 2.70 & 3.02 \\
5      & Inner Donut & 158    & 4.34 & 3.12 & 3.11 & 172    & 4.23 & 3.14 & 3.43 \\
5      & Outer Donut & 160    & 6.11 & 3.40 & 3.44 & 178    & 6.15 & 5.36 & 5.28 \\
5      & Edge        & 104    & 4.40 & 4.27 & 4.32 & 141    & 4.23 & 4.25 & 4.25 \\ \midrule
\multicolumn{2}{c|}{Mean}           & -         & 6.20 & \textbf{3.56} & \underline{3.75}  &  -     & 4.91 & \textbf{3.97} & \underline{4.10} \\ \bottomrule
\end{tabular}    
\end{table*}

\begin{table*}[]
    \centering
\caption{The testing RMSE (mV) of AC Scan $\Vmin$ prediction with 75\% data for training}
    \label{tab:main-rslt-ac-scan}
\begin{tabular}{@{}cl|cccc|cccc@{}}
\toprule
\multicolumn{2}{c|}{Temperature}& \multicolumn{4}{c|}{-45\degree C} & \multicolumn{4}{c}{25\degree C} \\ 
Wafer ID & Wafer Zone       & \# Die & Linear Regression  & Bias Alignment   & Restricted BA  & \# Die & Linear Regression  & Bias Alignment   & Restricted BA  \\ \midrule
1      & Center      & 166    & 7.71 & 5.08 & 5.01 & 164    & 5.49 & 4.72 & 4.62 \\
1      & Inner Donut & 162    & 6.70 & 6.33 & 6.49 & 159    & 6.54 & 6.12 & 6.20 \\
1      & Outer Donut & 158    & 5.29 & 5.53 & 5.45 & 173    & 8.90 & 5.79 & 5.63 \\
1      & Edge        & 90     & 6.66 & 5.85 & 6.19 & 123    & 9.43 & 8.32 & 8.44 \\ \midrule
2      & Center      & 152    & 12.06& 5.57 & 6.00 & 165    & 13.69& 5.17 & 6.10 \\
2      & Inner Donut & 153    & 7.35 & 6.77 & 6.83 & 170    & 7.23 & 6.23 & 5.81 \\
2      & Outer Donut & 155    & 6.10 & 6.34 & 6.58 & 170    & 6.16 & 6.29 & 6.41 \\
2      & Edge        & 93     & 5.27 & 4.61 & 4.48 & 121    & 7.66 & 6.19 & 5.88 \\ \midrule
3      & Center      & 54     & 11.52& 6.84 & 7.22 & 151    & 7.94 & 5.88 & 6.37 \\
3      & Inner Donut & 110    & 8.72 & 5.82 & 6.23 & 165    & 7.77 & 6.03 & 6.25 \\
3      & Outer Donut & 126    & 7.82 & 6.30 & 6.59 & 172    & 6.88 & 6.78 & 6.95 \\
3      & Edge        & 89     & 8.98 & 8.43 & 8.84 & 127    & 7.24 & 7.83 & 8.52 \\ \midrule
4      & Center      & 138    & 6.48 & 4.67 & 5.10 & 160    & 5.78 & 5.64 & 5.96 \\
4      & Inner Donut & 157    & 9.33 & 6.59 & 6.72 & 172    & 6.09 & 4.96 & 5.48 \\
4      & Outer Donut & 154    & 12.21& 7.18 & 7.30 & 168    & 11.23& 7.29 & 7.38 \\
4      & Edge        & 91     & 10.59& 6.52 & 6.65 & 123    & 9.07 & 8.15 & 8.34 \\ \midrule
5      & Center      & 153    & 5.54 & 5.38 & 5.01 & 172    & 6.53 & 6.18 & 6.17 \\
5      & Inner Donut & 158    & 6.09 & 6.31 & 6.24 & 170    & 5.86 & 5.66 & 5.78 \\
5      & Outer Donut & 159    & 6.95 & 6.56 & 6.83 & 176    & 9.29 & 6.22 & 6.19 \\
5      & Edge        & 103    & 6.39 & 6.44 & 6.89 & 136    & 7.30 & 6.87 & 6.20 \\ \midrule
mean   & -           & -         & 8.05 & \textbf{6.17} & \underline{6.33}  & -     & 8.04 & \textbf{6.31} & \underline{6.43} \\ \bottomrule
    \end{tabular}
\end{table*}

\begin{table*}[]
    \centering
\caption{The testing RMSE (mV) of MBST $\Vmin$ prediction with 75\% data for training}
    \label{tab:main-rslt-mbst}
\begin{tabular}{@{}cl|cccc|cccc@{}}
\toprule
\multicolumn{2}{c|}{Temperature}& \multicolumn{4}{c|}{-45\degree C} & \multicolumn{4}{c}{25\degree C} \\ 
Wafer ID & Wafer Zone       & \# Die & Linear Regression  & Bias Alignment   & Restricted BA  & \# Die & Linear Regression  & Bias Alignment   & Restricted BA  \\ \midrule
1      & Center      & 166    & 8.73  & 2.58 & 2.64 & 165    & 4.54  & 2.85 & 3.40 \\
1      & Inner Donut & 162    & 9.91  & 3.25 & 3.37 & 161    & 8.97  & 3.53 & 4.18 \\
1      & Outer Donut & 158    & 9.34  & 4.76 & 4.72 & 173    & 10.10 & 4.49 & 4.45 \\
1      & Edge        & 90     & 13.10 & 9.76 & 11.94& 120    & 15.98 & 9.77 & 11.17\\ \midrule
2      & Center      & 135    & 6.46  & 3.69 & 4.04 & 166    & 11.05 & 3.43 & 3.49 \\
2      & Inner Donut & 136    & 5.19  & 3.70 & 3.69 & 172    & 8.07  & 2.68 & 2.71 \\
2      & Outer Donut & 80     & 5.48  & 3.92 & 3.50 & 172    & 5.62  & 3.74 & 3.71 \\
2      & Edge        & 59     & 10.54 & 4.83 & 4.85 & 124    & 13.17 & 7.29 & 7.08 \\ \midrule
3      & Center      & 54     & 8.99  & 3.35 & 3.52 & 152    & 8.24  & 4.20 & 4.10 \\
3      & Inner Donut & 111    & 5.54  & 3.27 & 3.29 & 165    & 5.54  & 2.66 & 2.62 \\
3      & Outer Donut & 125    & 5.67  & 3.04 & 3.05 & 174    & 6.70  & 2.56 & 2.56 \\
3      & Edge        & 88     & 9.05  & 6.69 & 6.69 & 123    & 11.83 & 7.32 & 7.34 \\ \midrule
4      & Center      & 138    & 30.50 & 4.23 & 6.24 & 161    & 24.39 & 4.22 & 5.68 \\
4      & Inner Donut & 157    & 24.24 & 3.20 & 3.13 & 174    & 14.82 & 3.08 & 3.21 \\
4      & Outer Donut & 153    & 22.69 & 3.46 & 4.35 & 171    & 10.59 & 3.01 & 3.04 \\
4      & Edge        & 91     & 28.54 & 5.12 & 6.96 & 128    & 13.90 & 5.28 & 6.32 \\ \midrule
5      & Center      & 153    & 19.38 & 4.62 & 4.65 & 172    & 18.45 & 4.64 & 4.89 \\
5      & Inner Donut & 158    & 18.20 & 4.44 & 4.51 & 172    & 24.10 & 4.31 & 4.34 \\
5      & Outer Donut & 159    & 19.46 & 4.30 & 4.42 & 178    & 25.02 & 4.24 & 4.42 \\
5      & Edge        & 104    & 13.10 & 5.80 & 5.78 & 138    & 19.39 & 6.69 & 6.77 \\ \midrule
mean   & -           & -         & 16.25 & \textbf{4.45} & \underline{4.91}  & -     & 14.47 & \textbf{4.63} & \underline{4.94} \\ \bottomrule
    \end{tabular}
\end{table*}

\subsection{Restricted Bias Alignment (RBA) for $\Vmin$ Prediction}
RBA operates under the assumption that intra-wafer and inter-wafer variations are independent. This assumption is motivated by observations from \cref{fig:wafer-variation}, where the intra-wafer variation appears consistent across wafers, while the inter-wafer variation remains stable across wafer regions.

Moreover, RBA introduces the use of class probe features to model inter-wafer variations. By leveraging the pre-learned intra-wafer variations and the predicted inter-wafer variations, RBA demonstrates the capability to predict $\Vmin$ for a die from a new wafer without necessitating further data collection.

\subsubsection{Problem Formulation}
Denote $\z_i \in \R^{k}$ as the vector of $k$ class probe features of the $i$-th wafer, RBA models $\Vmin$ as

\begin{equation}
     \Y_{i,j} = \X_{i,j} \w + \bias_{i}(\z_i) + \bias_{j}^{intra} + \epsilon   
\end{equation}
where $\bias_{j}^{intra}$ accounts for the $\Vmin$ shift caused by intra-wafer variations; $\bias_{i}^{inter}(\z_i)$ is a linear model of $\z_i$, representing the voltage bias of wafer $i$:
\begin{equation}
    \bias_{i}^{inter} = \z_i \w_{\z} + \bias_{\z} + \epsilon_{\z}
\end{equation}

We construct a loss function $\loss_{RBA}$ as the sum of square residuals of the 
$\Vmin$ prediction across the whole training set:
\begin{equation}\label{eq:rba-loss}
    \loss_{RBA} := \ \sum_{i,j} ||\Y_{i,j}-\X_{i,j} \w - \bias_{i}^{inter} - \bias_{j}^{intra} ||_2^2
\end{equation}
where $\bm{\bias}$ is a set of $\Vmin$ shift of inter- and intra-wafer variations. We minimize the loss function it to estimate ${\w}$, $\bm{\bias}^{inter}$, and $\bm{\bias}^{intra}$:
\begin{equation}\label{eq:rba-obj}
    \widehat{\w}, \widehat{\bm{\bias}}^{inter}, \widehat{\bm{\bias}}^{intra} = \argmin_{\w,\bm{\bias}^{inter}, \bm{\bias}^{intra}}  \loss_{RBA}\left(\w, \bm{\bias}^{inter}, \bm{\bias}^{intra}\right)
\end{equation}

\subsubsection{Solution} To directly solve \cref{eq:rba-obj} is complicated. We adopt an alternative one-step gradient descent approach, integrating a novel initialization method to accelerate the convergence. 
Denote ${\w}_{(t)}$, $\bm{\bias}^{inter}_{(t)}$, and $\bm{\bias}^{intra}_{(t)}$ as the estimated parameters in the training step $t$. 

In step 0, we utilize BA to initialize parameters: ${\w}_{(0)}$ takes the value of the right-hand side of \cref{eq:ba-w}; $\bm{\bias}^{inter}_{(0)}$, and $\bm{\bias}^{intra}_{(0)}$ are the mean vector of the right-hand side of \cref{eq:ba-bias} by column and row, respectively.

In step $t$, we first optimize ${\w}_{(t)}$ to minimize $\loss_{RBA}$, condition on $\bm{\bias}^{inter}_{(t-1)}$ and and $\bm{\bias}^{intra}_{(t-1)}$:
\begin{equation}
    {\w}_{(t)} = \argmin_{\w} \loss_{RBA}\left(\w, \bm{\bias}^{inter}_{(t-1)}, \bm{\bias}^{intra}_{(t-1)}\right)
\end{equation}

This is a linear regression problem and the solution is 
\begin{align}\label{eq:rba-w-t}
    \w_{(t)} = \big({\X}^T {\X} \big) ^{-1} {\X}^T {\Y}_{(t)}
\end{align}
where $(\X, \Y_{(t)})$ is the concatenation of data of all wafer zones $(\X_{i,j}, \Y_{i,j}- \bias_{i;(t-1)}^{inter} - \bias_{j;(t-1)}^{intra})$.

Then, we adopt the chain rule to update biases:
\begin{align}\label{eq:rba-inter-bias-t}
    \bm{\bias}^{inter}_{(t)} = \bm{\bias}^{inter}_{(t-1)} - \eta \left(\frac{\partial \loss_{RBA}}{\partial \w_{(t)}} \cdot \frac{\partial \w_{(t)}}{\partial \bm{\bias}^{inter}_{(t-1)}}\right)^T \\ \label{eq:rba-intra-bias-t}
    \bm{\bias}^{intra}_{(t)} = \bm{\bias}^{intra}_{(t-1)} - \eta \left(\frac{\partial \loss_{RBA}}{\partial \w_{(t)}} \cdot \frac{\partial \w_{(t)}}{\partial \bm{\bias}^{intra}_{(t-1)}}\right)^T
\end{align}
where $\eta$ is a hyper-parameter of learning rate.

Once the training process is done, we optimize the coefficients of class probe features:
\begin{equation}
    \widehat{\w}_{\z}, \widehat{\bias}_{\z} = \argmin_{\w_{\z}, \bias_{\z}} \sum_{i} ||\widehat{\bias}^{inter}_{i} - \z_i \w_{\z} - \bias_{\z}||_2^2
\end{equation}

\subsubsection{Discussion} RBA effectively separates the influence of inter- and intra-wafer variations on $\Vmin$ shift. This decoupling mechanism distinguishes RBA from BA, mitigating potential overfitting concerns particularly when dealing with small training datasets, thereby bolstering overall data efficiency.

For a testing die $(\x^{test}_{i,j}, \y^{test}_{i,j})$ form the $j$-th zone of the $i$-th wafer, the $\Vmin$ prediction of RBA is 
\begin{equation}
    \widehat{\y}^{test} = \x^{test} \widehat{\w} + \widehat{\bias}_{i}^{inter}(\z_i) + \widehat{\bias}_{j}^{intra}
\end{equation}
where
\begin{equation}
    \widehat{\bias}_{i}^{inter}(\z_i) = \z_i \w_{\z} +\bias_{\z}
\end{equation}

By incorporating class probe features to capture inter-wafer variations, RBA possesses the capability for deployment in $\Vmin$ prediction without necessitating re-training or measuring $\Vmin$ for any dies from a new wafer. This feature enhances the practical applicability and efficiency of RBA in product testing scenarios.

\section{Experimental Results}
We conduct experiments to demonstrate the efficacy of our approach RBA for addressing inter- and intra-wafer variations on thousands of 16nm automotive chips. We aim to illustrate: 1) the effectiveness of $\Vmin$ bias alignment, 2) the data efficiency and robustness of RBA, 
3) the capability of class probe features to capture inter-wafer variation, and 4) the ability of RBA to predict $\Vmin$ of dies from a new wafer.

\paragraph{Description of Data Collection} We get the class probe data of each wafer from the foundry. During the testing flow of product manufacturing, $\Vmin$, including DC Scan $\Vmin$, AC Scan $\Vmin$, and MBIST $\Vmin$ are measured at at three different temperatures: -45\degree C (cold), 25\degree C (room), and 125\degree C (hot). Similarly, parametric test and POSt test data were collected under different temperatures from Automatic Test Equipment (ATE) testers.

Our dataset has several wafers. Each wafer is partitioned into 4 zones: center, inner donut, outer donut, and edge. The visualization of this partition is shown in \cref{fig:wafer-region}, and the number of dies in each wafer zone is listed in \cref{tab:wafer-zone}. Due to the expensive cost of $\Vmin$ test, only a subset of dies is performed the $\Vmin$ test for a certain test pattern.

\paragraph{RBA Settings} RBA leverages 5 parametric test features and 2 class probe features as input to predict $\Vmin$. All features are selected by the Correlation Feature Selection algorithm \cite{CFS}, and pass the causation check by our testing engineer. The hyper-parameter learning rate $\eta$ is set to 0.1. We terminate the training process when the relative improvement of the loss function $\loss_{RBA}$ is smaller than 0.001.

\paragraph{Baseline Settings} We compare RBA with 2 baselines: linear regression, and the proposed Bias Alignment (BA). Linear regression is trained over all of the label data without handling process variations. To ensure a fair comparison, all the baselines and RBA share the same general configurations, including input features, training data, and testing data.

\begin{table}[]
    \centering
\caption{Inter-wafer $\Vmin$ shift (mV) from wafer 1 estimated by RBA}
\label{tab:inter-vmin-shift}
\begin{tabular}{@{}lcccc@{}}
\toprule
Wafer ID & Temp. & DC Scan $\Vmin$ & AC Scan $\Vmin$ & MBST $\Vmin$   \\ \midrule
2        & -45\degree C   & -18.78  & -6.93   & -14.41 \\
3        & -45\degree C   & -4.32   & -4.45   & -6.87  \\
4        & -45\degree C   & -5.98   & 5.69    & -28.93 \\
5        & -45\degree C   & -3.28   & 2.35    & 17.68  \\\midrule
2        & 25\degree C    & -9.91   & -10.03  & -18.75 \\
3        & 25\degree C    & 2.18    & -4.10    & -8.77  \\
4        & 25\degree C    & 0.62    & 6.60     & -27.99 \\
5        & 25\degree C    & -0.37   & 7.48    & -18.02 \\ \bottomrule
\end{tabular}
\end{table}

\begin{table}[]
\centering
\caption{Intra-wafer $\Vmin$ shift (mV) from center zone estimated by RBA}
\label{tab:intra-vmin-shift}
\begin{tabular}{@{}lcccc@{}}
\toprule
Wafer Zone    & Temp. & DC Scan $\Vmin$ & AC Scan $\Vmin$ & MBST $\Vmin$   \\ \midrule
Inner Donut   & -45\degree C   & 2.48    & 4.16    & 7.35   \\
Outer Donut   & -45\degree C   & 2.87    & 6.81    & -2.17  \\
Edge          & -45\degree C   & 1.36    & 5.65    & -1.18  \\\midrule
Inner Donut   & 25\degree C    & 2.32    & 4.22    & -1.03  \\
Outer Donut   & 25\degree C    & 3.13    & 6.95    & -2.24  \\
Edge          & 25\degree C    & -0.35   & 3.47    & -11.15 \\ \bottomrule
\end{tabular}
\end{table}

\subsection{Effectiveness of $\Vmin$ bias alignment} \label{sec:rslt-main}
We aim to showcase the effectiveness of $\Vmin$ bias alignment in the $\Vmin$ prediction task.

\paragraph{Experimental Settings} 
We consider all three types of $\Vmin$ (DC Scan, AC Scan, and MBST), tested at cold and room temperatures. Our dataset comprises 5 wafers tested under these conditions. For each wafer zone, we allocate 75\% of the dies for training and the remaining 25\% for testing. The methods under consideration include linear regression, BA, and RBA.

\paragraph{Results} We report the testing RMSE of $\Vmin$ prediction in each wafer zone, and the average result across the whole testing dataset in \cref{tab:main-rslt-dc-scan} for DC Scan $\Vmin$, \cref{tab:main-rslt-ac-scan} for AC Scan $\Vmin$, and \cref{tab:main-rslt-mbst} for MBST $\Vmin$. The bold number represents the best method, and the underlined number represents the second-best method. In each $\Vmin$ prediction task, both BA and RBA consistently outperform the baseline linear regression. This suggests that the bias alignment technique effectively captures process variations. Notably, the performance gap between BA and RBA is minimal, indicating a weak dependency between inter- and intra-wafer variations.

Additionally, we show the inter- and intra-wafer $\Vmin$ shift estimated by our approach RBA in \cref{tab:inter-vmin-shift} and \cref{tab:intra-vmin-shift}, respectively. The variance of both types of $\Vmin$ shift is substantial and cannot be disregarded. A significant $\Vmin$ shift notably impacts the accuracy of linear regression. For instance, linear regression performs badly for predicting DC Scan $\Vmin$ of wafer 2 at the cold temperature, where a -18.78mV $\Vmin$ shift is estimated by RBA.

\begin{table}[]
    \centering
    \caption{RBA with different fractions of data for training}
    \label{tab:rba-efficiency}
\begin{tabular}{@{}lcc|cc@{}}
\toprule
Temperature        & \multicolumn{2}{c|}{-45\degree C}   & \multicolumn{2}{c}{25\degree C}    \\
Training data frac. & 75\%  & 5\%   & 75\%  & 5\%   \\ \midrule
DC Scan $\Vmin$ (mV) & 3.75  & 3.85  & 4.10  & 3.82  \\
AC Scan $\Vmin$ (mV) & 6.33  & 6.42  & 6.43  & 6.54  \\
MBST $\Vmin$ (mV)   & 4.91  & 5.03  & 4.94  & 5.18  \\ \bottomrule
\end{tabular}
\end{table}

\subsection{Data Efficiency and Robustness of RBA}
We present the performance of RBA on small training datasets to demonstrate its data efficiency and robustness.
\paragraph{Experimental Settings} We split 5\% dies in each wafer zone for training, and the rest for testing. All other configurations are the same as those in \cref{sec:rslt-main}.

\paragraph{Results} The $\Vmin$ prediction accuracy of RBA is listed in \cref{tab:rba-efficiency}. While the fraction of training data is reduced from 75\% to 5\% (around 7 dies in each wafer zone), RBA's accuracy is stable, indicating its superior data efficiency and robustness.

\begin{table}[]
    \centering
\caption{Top 1 linear correlation between class probe features and inter-waver $\Vmin$ shift estimated by RBA}
\label{tab:rba-class-probe}
\begin{tabular}{@{}lccc@{}}
\toprule
Test Pattern               & DC Scan $\Vmin$ & AC Scan $\Vmin$ & MBST $\Vmin$ \\ \midrule
Top 1 Linear Corr.  & 0.892           & 0.948           & 0.953        \\ \bottomrule
\end{tabular}
\end{table}

\begin{table}[]
    \centering
\caption{Coefficient of determination of the linear model using 2 class probe features to predict $\Vmin$ shift estimated by RBA}
\label{tab:rba-class-probe-r2}
\begin{tabular}{@{}lccc@{}}
\toprule
Test Pattern               & DC Scan $\Vmin$ & AC Scan $\Vmin$ & MBST $\Vmin$ \\ \midrule
$R^2$  & 0.509           & 0.631           & 0.792        \\ \bottomrule
\end{tabular}
\end{table}

\subsection{Class Probe Features Capturing Inter-Wafer Variation}\label{sec:rslt-rba-probe}
We demonstrate that the dependency between class probe features and inter-wafer variations estimated by RBA is really high, indicating the motivation to leverage wafer-level class probe features to model wafer-to-wafer variation is plausible.
\paragraph{Experimental Settings} Our dataset has 10 wafers whose DC Scan, AC Scan, and MBST $\Vmin$ are tested at the hot temperature. In the first step, we employ RBA on these wafers to collect 9 $\Vmin$ shift terms relative to a base wafer. Subsequently, we correlate these shifts with each class probe feature, reporting the highest absolute value of the Pearson correlation coefficient. A higher coefficient indicates a stronger linear correlation. In the second step, we utilize the $\Vmin$ shifts of 6 wafers to fit a linear model for 2 class probe features and evaluate its testing coefficient of determination ($R^2$) on the remaining 4 wafers.

\paragraph{Results} In \cref{tab:rba-class-probe}, it is evident that for each $\Vmin$ test pattern, there exists a class probe feature with a correlation coefficient of at least 0.89, indicating (1) the credibility of the inter-wafer $\Vmin$ shift estimated by RBA, and (2) the informativeness of class probe data in modeling $\Vmin$ shift across wafers.

\cref{tab:rba-class-probe-r2} reports the test accuracy of using class probe features to model $\Vmin$ shift. The $R^2$ score of each $\Vmin$ test pattern is proportion to the Pearson score in \cref{tab:rba-class-probe}. While MBST $\Vmin$ shift predictors appear promising, we encounter difficulty in obtaining a sufficiently accurate predictor for the DC/AC Scan $\Vmin$ shift. This challenge may stem from the small size of the training dataset, leading to an increased variance. We defer this issue to future research endeavors where a larger pool of tested wafers can be obtained.

\begin{table}[]
    \centering
\caption{The testing RMSE (mV) of $\Vmin$ tested at 125\degree C}
\label{tab:rba-cfinal}
\begin{tabular}{@{}lccc@{}}
\toprule
Method               & DC Scan $\Vmin$ & AC Scan $\Vmin$ & MBST $\Vmin$ \\ \midrule
Linear Regression  & \textbf{7.33}           & 9.40           & 14.24        \\ 
RBA & 8.07  & \textbf{8.98} & \textbf{9.18}\\
\bottomrule
\end{tabular}
\end{table}

\subsection{RBA for $\Vmin$ Prediction of New Wafer}
We assess the effectiveness of RBA in predicting $\Vmin$ of new wafers, focusing on addressing inter- and intra-wafer variations.

\paragraph{Experimental Settings} We evaluate RBA on the $\Vmin$ prediction task, where $\Vmin$ is tested at 125\degree C. Following \cref{sec:rslt-rba-probe}, we use 6 wafers for training and 4 wafers for testing. The baseline model is linear regression.

\paragraph{Results} 
\cref{tab:rba-cfinal} presents the RMSE of RBA and linear regression. Owing to the limited number of wafers available in our dataset for training the inter-wafer $\Vmin$ shift predictor, RBA and linear regression yield comparable results for DC/AC Scan $\Vmin$ prediction. However, RBA exhibits a significant performance advantage over linear regression in the MBST $\Vmin$ prediction task, highlighting its efficacy in addressing inter- and intra-wafer variations.

\section{Conclusion}
This paper introduces restricted bias alignment (RBA), a $\Vmin$ prediction framework designed to systematically capture process variations in semiconductor manufacturing. By leveraging class probe features to model inter-wafer variations and utilizing parametric features to estimate intra-wafer variations, RBA offers a comprehensive approach to address the challenges posed by process variations.

Our experiments conducted on an industrial dataset demonstrate the effectiveness of RBA in mitigating the impact of process variations on $\Vmin$ prediction. The results highlight the practical utility and robustness of RBA in real-world semiconductor manufacturing scenarios, underscoring its potential to enhance manufacturing efficiency and reliability.

\section*{Acknowledgment}
The content of this paper has been developed with the support of Grant No. 1956313 from the National Science Foundation (NSF) and has also received partial funding from a Long Term University (LTU) grant provided by NXP.

\bibliographystyle{IEEEtran}
\bibliography{IEEEabrv, ref}

\begin{thebibliography}{10}
\providecommand{\url}[1]{#1}
\csname url@samestyle\endcsname
\providecommand{\newblock}{\relax}
\providecommand{\bibinfo}[2]{#2}
\providecommand{\BIBentrySTDinterwordspacing}{\spaceskip=0pt\relax}
\providecommand{\BIBentryALTinterwordstretchfactor}{4}
\providecommand{\BIBentryALTinterwordspacing}{\spaceskip=\fontdimen2\font plus
\BIBentryALTinterwordstretchfactor\fontdimen3\font minus \fontdimen4\font\relax}
\providecommand{\BIBforeignlanguage}[2]{{%
\expandafter\ifx\csname l@#1\endcsname\relax
\typeout{** WARNING: IEEEtran.bst: No hyphenation pattern has been}%
\typeout{** loaded for the language `#1'. Using the pattern for}%
\typeout{** the default language instead.}%
\else
\language=\csname l@#1\endcsname
\fi
#2}}
\providecommand{\BIBdecl}{\relax}
\BIBdecl

\bibitem{VminBin}
W.-C. Lin, C.~Chen, C.-H. Hsieh, J.~C.-M. Li, E.~J.-W. Fang, and S.~S.-Y. Hsueh, ``Ml-assisted vminbinning with multiple guard bands for low power consumption,'' in \emph{2022 IEEE International Test Conference (ITC)}, 2022, pp. 213--218.

\bibitem{VminTest}
C.~He and Y.~Yu, ``Wafer level stress: Enabling zero defect quality for automotive microcontrollers without package burn-in,'' in \emph{2020 IEEE International Test Conference (ITC)}, 2020, pp. 1--10.

\bibitem{Odometer}
J.~Keane, W.~Zhang, and C.~H. Kim, ``An array-based odometer system for statistically significant circuit aging characterization,'' \emph{IEEE Journal of Solid-State Circuits}, vol.~46, no.~10, pp. 2374--2385, 2011.

\bibitem{RODesign}
T.-B. Chan, P.~Gupta, A.~B. Kahng, and L.~Lai, ``Ddro: A novel performance monitoring methodology based on design-dependent ring oscillators,'' in \emph{Thirteenth International Symposium on Quality Electronic Design (ISQED)}, 2012, pp. 633--640.

\bibitem{Accumulative}
Y.-T. Kuo, W.-C. Lin, C.~Chen, C.-H. Hsieh, J.~C.-M. Li, E.~Jia-Wei~Fang, and S.~S.-Y. Hsueh, ``Minimum operating voltage prediction in production test using accumulative learning,'' in \emph{2021 IEEE International Test Conference (ITC)}, 2021, pp. 47--52.

\bibitem{SVMFmaxBinning}
Q.~Shi, X.~Wang, L.~Winemberg, and M.~M. Tehranipoor, ``On-chip sensor selection for effective speed-binning,'' \emph{Analog Integrated Circuits and Signal Processing}, vol.~88, pp. 369--382, 2016.

\bibitem{GPModel}
J.~Chen, J.~Zeng, L.-C. Wang, M.~Mateja, and J.~Rearick, ``Predicting multi-core system fmax by data-learning methodology,'' in \emph{Proceedings of 2010 International Symposium on VLSI Design, Automation and Test}, 2010, pp. 220--223.

\bibitem{ML-Assisted}
W.-C. Lin, C.~Chen, C.-H. Hsieh, J.~C.-M. Li, E.~J.-W. Fang, and S.~S.-Y. Hsueh, ``Ml-assisted vminbinning with multiple guard bands for low power consumption,'' in \emph{2022 IEEE International Test Conference (ITC)}, 2022, pp. 213--218.

\bibitem{yin-mln-itc}
Y.~Yin, R.~Chen, C.~He, and P.~Li, ``Domain-specific machine learning based minimum operating voltage prediction using on-chip monitor data,'' in \emph{2023 IEEE International Test Conference (ITC)}, 2023, pp. 99--104.

\bibitem{CFS}
M.~A. Hall, ``Correlation-based feature selection for machine learning,'' Ph.D. dissertation, The University of Waikato, 1999.

\end{thebibliography}

\end{document}